\documentclass[a4paper]{article}

\usepackage[english]{babel}
\usepackage[utf8x]{inputenc}
\usepackage[T1]{fontenc}
\usepackage{multirow}
\usepackage{authblk}
\usepackage[a4paper,top=3cm,bottom=2cm,left=3cm,right=3cm,marginparwidth=1.75cm]{geometry}

\usepackage{amsmath}
\usepackage{graphicx}
\usepackage[colorinlistoftodos]{todonotes}
\usepackage[colorlinks=true, allcolors=blue]{hyperref}
\usepackage{subcaption}

\title{Fully-automated deep learning slice-based muscle estimation from CT images for sarcopenia assessment}

\author[1]{Fahdi Kanavati}
\author[1]{Shah Islam}
\author{Zohaib Arain}
\author[1]{Eric O. Aboagye}
\author[1,2]{Andrea Rockall}
\affil[1]{Comprehensive Cancer Imaging Centre, Hammersmith Hospital, Imperial College London, UK}
\affil[2]{Dept of Radiology, Imperial College Healthcare NHS Trust}

\date{November, 2019}

\begin{document}
\maketitle

\section{Abstract}

\paragraph{Objective}
To demonstrate the effectiveness of using a deep learning-based approach for a fully automated slice-based measurement of muscle mass for assessing sarcopenia on CT scans of the abdomen without any case exclusion criteria.

\paragraph{Materials and Methods}
This retrospective study was conducted using a collection of public and privately available CT images (n = 1070). The method consisted of two stages: slice detection from a CT volume and single-slice CT segmentation. Both stages used Fully Convolutional Neural Networks (FCNN) and were based on a UNet-like architecture. Input data consisted of CT volumes with a variety of fields of view.   The output consisted of a segmented muscle mass on a CT slice at the level of L3 vertebra. The muscle mass is segmented into erector spinae, psoas, and rectus abdominus muscle groups. The output was tested against manual ground-truth segmentation by an expert annotator.

\paragraph{Results}
3-fold cross validation was used to evaluate the proposed method. The slice detection cross validation error was 1.41±5.02 (in slices). The segmentation cross validation Dice overlaps were 0.97±0.02, 0.95±0.04, 0.94±0.04 for erector spinae, psoas, and rectus abdominus, respectively, and 0.96±0.02 for the combined muscle mass.

\paragraph{Conclusion}
A deep learning approach to detect CT slices and segment muscle mass to perform slice-based analysis of sarcopenia is an effective and promising approach. The use of FCNN to accurately and efficiently detect a slice in CT volumes with a variety of fields of view, occlusions, and slice thicknesses was demonstrated.

\section{Summary}
Automated slice detection and muscle mass segmentation from CT data is efficient and feasible using deep learning.

\section{Key Points}

\begin{itemize}
    \item For slice detection, 3-fold cross validation mean slice error was 1.41±5.02 (median 0.50). The inter-radiologist error was 1.94±2.36 (median 0.80).
    \item For muscle segmentation, 3-fold cross validation Dice scores were 0.97±0.02, 0.95±0.04, 0.94±0.04 for erector spinae, psoas, and rectus abdominus, respectively, and 0.96±0.02 for the combined muscle mass.
    % \item Using an average of multiple slices is more robust compared to using a single slice.
\end{itemize}

\section{Introduction}

Single-slice CT analysis has garnered significant clinical interest in the past few years in regards to assessing sarcopenia as a surrogate marker of frailty \cite{shachar2016prognostic}.
Sarcopenia refers to loss of muscle mass and strength and is particularly relevant in oncology where it is typically found to be associated with poor outcomes  \cite{tegels2015sarcopenia,mir2012sarcopenia, kazemi2016computed}.
A slice extracted at the L3 vertebra is commonly chosen as a standard landmark by a majority of medical researchers for sarcopenia measurement \cite{shachar2016prognostic},  as muscle and adipose fat areas at L3 have been found to be the most linearly correlated to their whole-body counterparts \cite{kazemi2016computed}.

The main motivation for automating the process of single-slice CT analysis is to (a) accelerate research for studies that extract single-slice sarcopenia-based measurements, and (b) use it as part of a potential opportunistic screening tool that provides prognostic information to clinicians, based on clinically validated measurements, in cancer populations alongside the standardised evaluation of CT images at the time of initial staging.  Identification of sarcopenia, indicating frailty,  would allow appropriate stratification of patients to treatment regimes, including pre-habilitation and re-habilitation.
The current work-flow for single slice sarcopenia measurement is as follows: manual extraction of the L3 slice; this involves scrolling through the 3D image slice by slice until the L3 slice is found. Semi-automated segmentation software (e.g. Slice-O-Matic or ImageJ), which involves manual refinement, is then used to segment the skeletal muscle and adipose fat tissue. This process takes approximately 10 minutes per image, and it becomes time-consuming to run on large datasets.  In addition, due to the manual step, this has not been routinely incorporated into the daily workflow. 
In this paper, we propose an efficient deep learning method based on FCNN to detect the L3 slice and segment the skeletal muscle at that slice into Erector Spinae, Psoas, and Rectus Abdominus. A FCNN-based method was motivated by previous work \cite{tompson2015efficient, pfister2015flowing, payer2016regressing, cao2017realtime} that used FCNNs to predict confidence maps for landmark localisation.

\section{Materials and Methods}

Our method is composed of two stages, each of which is independently tested: 1) slice detection and 2) muscle segmentation. As a whole, the method takes as input a CT volume and outputs segmentations for the detected L3 slice.

\subsection{Dataset}

We collected, without any selection criteria, a diverse dataset consisting of CT images of cancer patients from multiple sources.  We did not exclude any cases due to metal implants. Two sets were obtained from the Cancer Imaging Archive (TCIA) \cite{clark2013cancer}: head and neck\cite{tcia_hnc} (n = 90), ovarian \cite{tcia_ovarian}, and colon \cite{tcia_colon, johnson2008accuracy} (n=674); a liver tumour dataset (n=96) was obtained from the LiTS segmentation challenge\cite{bilic2019liver}.
In addition, we used an ovarian cancer dataset (n=210) obtained from Hammersmith Hospital (HH), London. Ethical approval for retrospective analysis of the HH data was obtained under the Hammersmith and Queen Charlotte’s \& Chelsea Research Ethics Committee approval 05/QO406/178, and informed consent was waived, typical for retrospective analysis of anonymized imaging data. This resulted in a total of 1070 CT volumes. For slice detection, all were manually annotated. For segmentation, in total 1156 slices from the 1070 volumes were extracted from around the L3 vertebra, with multiple adjacent slices extracted from some of the volumes, and they were manually segmented.
 
 \subsection{Slice detection dataset preparation}

All the input 3D CT volumes were first converted into 2D Maximal Intensity Projection (MIP) images along the frontal and sagittal views, similarly to \cite{belharbi2017spotting};
 however, we computed a restricted MIP for the sagittal view in order to eliminate the outer edges of the pelvis and to have a clear view of the sacrum vertebrae, which is an essential reference point for determining the position of L3 if the annotator counts the vertebrae bottom-up. As the spine tends to be situated in the middle of the images in the majority of cases, we only computed the restricted sagittal MIP using the range [-20, 20] from the centre of the image; however, for the rare cases where the spine is not at the centre of the image, it is potentially possible to use a more elaborate image processing technique to attempt to detect the centre-line of the spine and centre it. As the CT images in the collected datasets had different slice thicknesses (0.5 to 7mm), we normalised the pixel size of the resulting MIPs to $1\times1mm^2$ to allow consistent input to the slice detection algorithm.

\begin{figure}[h]
    \centering
       \begin{subfigure}[b]{0.40\textwidth}
        \includegraphics[width=1\textwidth]{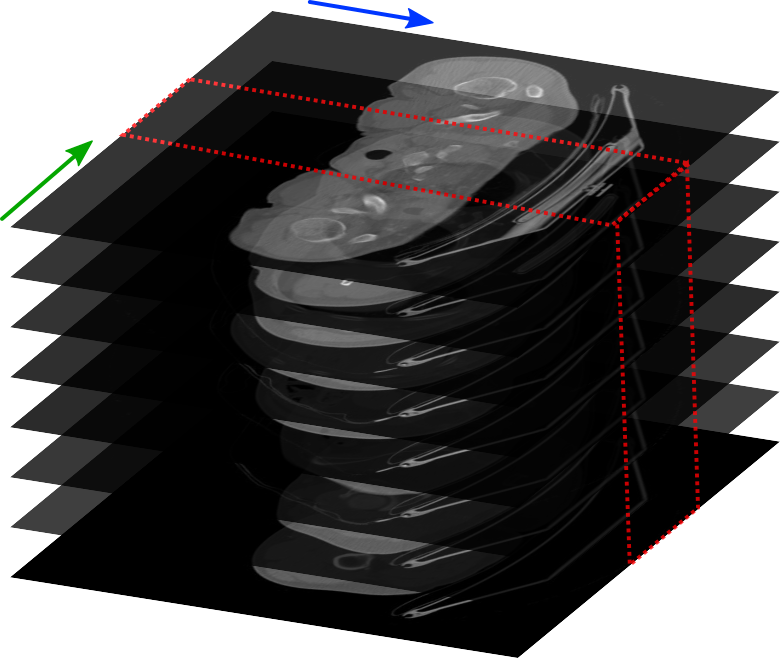}
\caption{A stack of CT slices.}
     
    \end{subfigure}
    
        \begin{subfigure}[t]{0.31\textwidth}
        \includegraphics[width=1\textwidth]{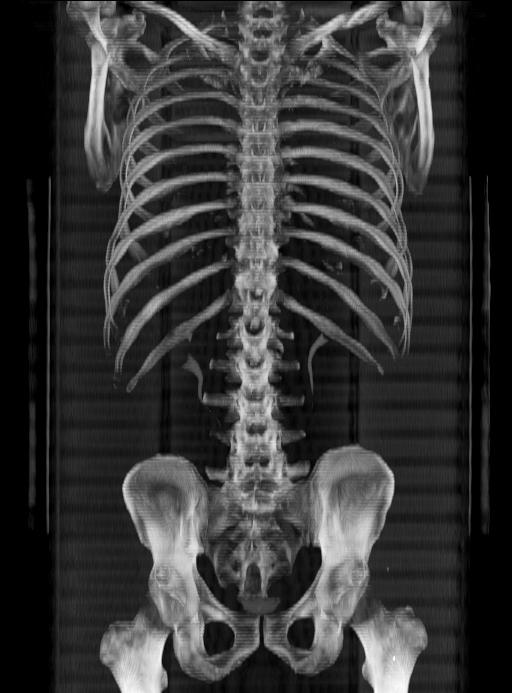}
\caption{Frontal MIP image obtained from performing a maximal intensity projection along the direction perpendicular to the frontal plane (blue in (a)).}
      
    \end{subfigure}
    \begin{subfigure}[t]{0.31\textwidth}
        \includegraphics[width=1\textwidth]{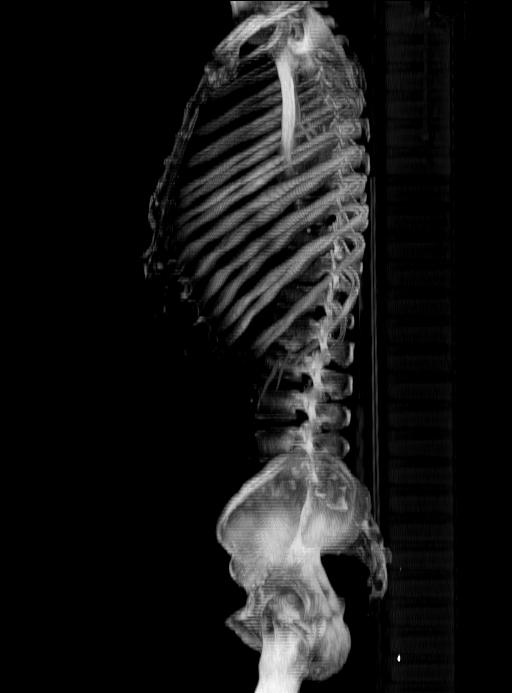}
\caption{Sagittal MIP image obtained from performing a maximal intensity projection along the direction perpendicular to the sagittal plane (green in (a)). Note that the vertebral bodies are not clearly visible.}
       
    \end{subfigure}
    ~ 
    \begin{subfigure}[t]{0.31\textwidth}
       \includegraphics[width=1\textwidth]{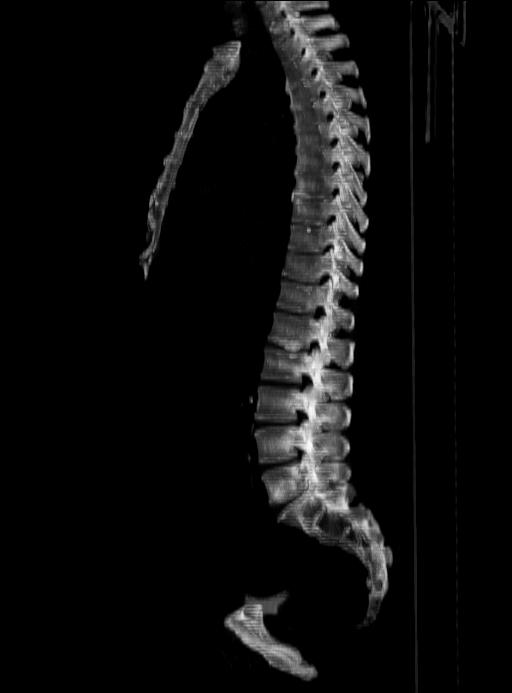}
\caption{Restricted sagittal MIP image computed only on the restricted range (in red) of the CT in (a).}

    \end{subfigure}

    \caption{Maximal intensity projection (MIP) images obtained from CT volumes.\label{fig:mip}}
\end{figure}

Finally, we thresholded the images between 100 HU and 1500 HU in order to eliminate the majority of soft tissue at the lower end of HU and minimise the effect of metal implants and artifacts above 1500 HU. The images were then mapped to 8bit ($[-127,127]$). Figure \ref{fig:mip} shows an example of MIP images obtained from a 3D CT volume.

The MIP images were annotated by two annotators: a radiologist with eight years of experience and an annotator with five years of experience working with CT images. For each image set, the annotator was presented with the frontal and restricted sagittal MIPs side by side, and the annotator clicked on the location of the L3 slice. The main landmark was chosen as the middle of the vertebra, lining up with the top edge of the transverse process. Only the position along the y-axis was recorded. Figure \ref{fig:image_overlays} shows an example of L3 slice annotations. It took about 2-4 seconds to manually annotate a single image. Nine images had disagreements between the annotators and 57 were ambiguous as there was uncertainty in assigning the location of the L3 vertebra. Further inspection by a senior radiologist with 18 years of experience as a consultant revealed that the majority of ambiguous cases consisted of patients with congenital vertebral anomalies, with the principal anomaly being transitional vertebra. Transitional vertebrae are ones that exhibit ambiguous characteristics; they are relatively common in the population (15-35\%)  \cite{carrino2011effect, uccar2013retrospective}, and they occur at the junction between spinal segments with various degrees of apparent transition: atlanto-occipital junction, cervicothoracic junction (with a cervical rib from C7), thoracolumbar junction (with lumbar rib at L1 or a 13th rib from T13), and lumbosacral junction (commonly referred to as Lumbosacral transitional vertebrae (LSTV)). Inaccurate identification of the correct level due to LSTV has led to procedures being carried out at the wrong vertebra level \cite{konin2010lumbosacral}.
Correct identification of L3 in ambiguous cases can only be resolved if the image contains a view of the whole spine  \cite{bron2007clinical, carrino2011effect}.
Transitional vertebra cases (57) were excluded, leaving 1006 images for the training process; however, we still evaluated the detection algorithm on the transitional cases to verify the output, as in real-world scenarios, such cases are expected to be encountered. The average (rounded down) L3 slice location from the two annotators is used as ground truth for training.

\subsection{Muscle segmentation dataset preparation}

\begin{figure}[h]
\centering
\includegraphics[width=1\textwidth]{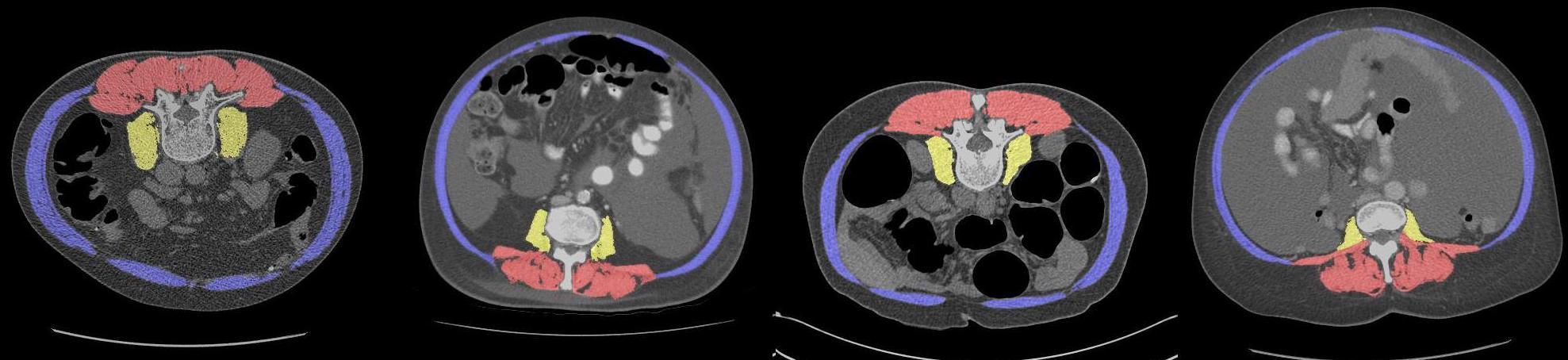}
 
\caption{\label{fig:example_seg} Examples of CT slices with manual segmentations of the muscle into erector spinae, psoas, and rectus abdominus. The colon dataset contained a mix of prone and supine patients.   }
\end{figure}

\begin{figure}
    \centering
                \begin{subfigure}[b]{0.23\textwidth}
        \includegraphics[width=1\textwidth]{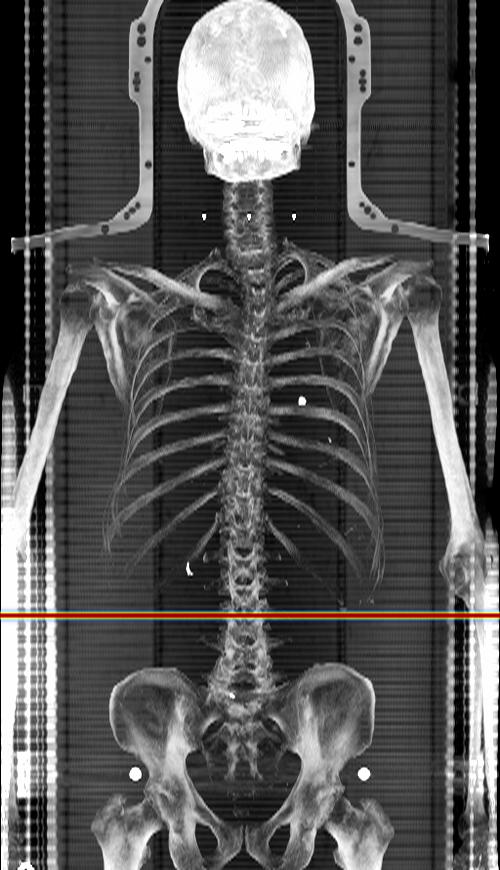}
        \caption{}
    \end{subfigure}
    \begin{subfigure}[b]{0.23\textwidth}
        \includegraphics[width=1\textwidth]{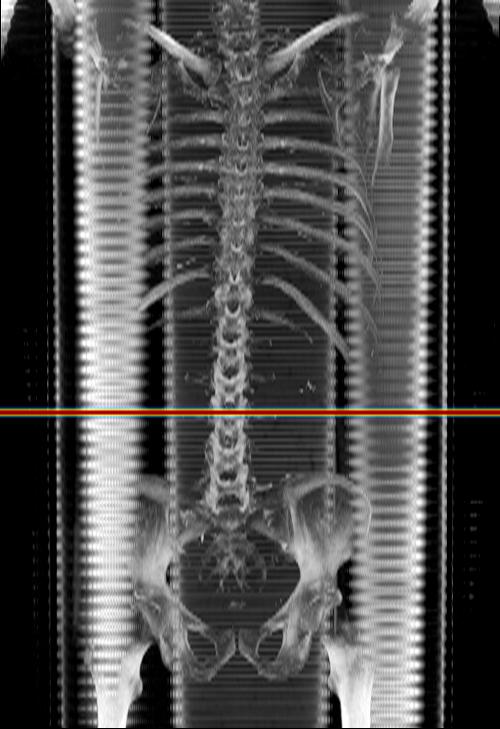}
        \caption{}
         \end{subfigure}
            \begin{subfigure}[b]{0.23\textwidth}
        \includegraphics[width=1\textwidth]{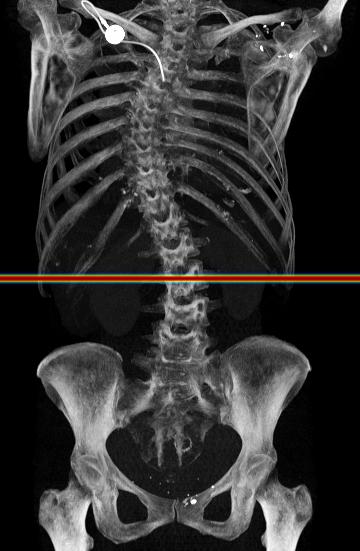}
        \caption{}
    \end{subfigure}
      \begin{subfigure}[b]{0.23\textwidth}
        \includegraphics[width=1\textwidth]{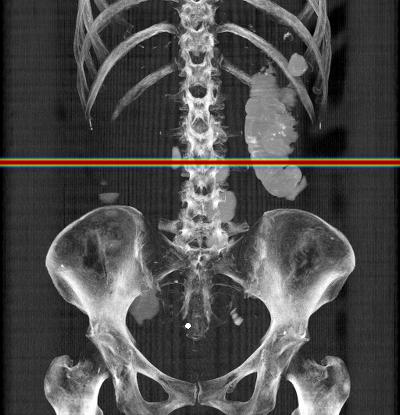}
        \caption{}
         \end{subfigure}
\caption{\label{fig:image_overlays}Examples (a-d) of different images with the ground-truth 1D  confidence maps for the location of the  L3  vertebra overlaid on top of the images. The dataset consisted of different images with a variety of fields of view, slice thicknesses, and artefacts.}
\end{figure}

A total of 1156 transverse slices were extracted from the 1070 volumes from around the L3 vertebra, with multiple adjacent slices extracted from some volumes with small slice thickness. For the majority of the images, the manual segmentation was carried out in ITKsnap \cite{yushkevich2006user}  by a single radiologist with eight years of experience; for efficiency while manually segmenting, the extracted slices were stacked up as a single volume to allow loading them all at once within the viewer. The slices were segmented into erector spinae, psoas, and rectus abdominus. Figure \ref{fig:example_seg} shows examples of the segmentations.
We clipped the image intensities to the $[-250HU, 250HU]$, so as to allow enough extended range to include the muscle ($[-29HU, 150HU]$). We then normalised the images to the $[-1.0, 1.0]$ range. All the images were of a fixed size of $512\times512$ pixels.
On a subset of the head and neck dataset (n=64), an annotator with 3 years of experience of human anatomy performed an independent slice selection at L3 and muscle segmentation. These segmentations were then checked by a senior radiologist.  This set was used for evaluation of inter-rater agreement between human annotators.

\subsection{Deep learning model}

For both the slice detection and segmentation, we used FCNNs that are variants of the UNet architecture. The UNet architecture  \cite{ronneberger2015u} consists of multiple down-sampling (encoder path) and up-sampling (decoder path) blocks, with the latter mirroring the former and with skip connections from the encoder to the decoder path. Each block consists of a set of convolutional units, where each unit is a sequence convolution, batch normalisation, and non-linear activation.
For slice detection, we modified the UNet architecture to output 1D heatmaps. To do so, we introduced a global horizontal max-pooling that reduces the dimension from 2D to 1D by taking the maximum value only along the horizontal line. 

\subsection{Augmentation}

We employed image augmentation to help artificially increase the number of images that the CNN encounters during training. Augmentation typically helps in improving generalisation performance. Image transformations were applied to the input images create artificial variants. We used a set of image transformations such as: horizontal flipping, scaling, intensity offsets, and piece-wise affine deformation. For slice detection, we additionally used region drop-outs and over-exposure (to simulate occlusions), and vertical image sub-sampling (to simulate different slice thicknesses). In MIP images, occlusions can show up due to the presence of metal implants, bowel content, or contrast agents; region drop-outs and over-exposures can help make the algorithm less susceptible to such occlusions. To simulate images with a variety of slice thicknesses (up to $7mm$), images were randomly down-sampled along the vertical axis and then up-sampled back to their original size using linear interpolation.

\subsection{Data analysis}
To assess the performance of the slice detection and segmentation, 3-fold cross validation was performed. The slice detection error was measured using the absolute difference between the ground truth position of L3 and the predicted position. The muscle segmentation accuracy was measured using the Dice score, which is defined as twice the area of the overlap between the predicted segmentation and the ground truth segmentation divided by the sum of both areas. The mean and standard deviation from the cross validation were computed.
Sarcopenia assessment is typically done by computing the following: Skeletal Muscle Index (SMI), which is defined as the skeletal muscle area (centimeters squared) divided by patient height (meters squares), and Muscle Attenuation (MA), the average intensity in Hounsfield Units (HU) within the muscle area. As no patient height data was available, we only report the muscle area as a proxy measure for SMI.

\section{Results}

\subsection{Slice detection}

We evaluated the slice detection method using either the frontal or restricted sagittal as input to the CNN. Table \ref{tab:crossval_error} summarises the results of 3-fold cross validation for slice detection. The absolute error was 1.53±4.22 slices (2±4.56mm).

We additionally applied the slice detection on the excluded set of transitional vertebra cases.
Outputs in all cases consisted in one of the two adjacent true candidate vertebrae or, in some cases, both. 
Figure \ref{fig:example_output} shows example slice detection prediction outputs and Fig. \ref{fig:transitional} shows a slice detection output on a transitional vertebra case.

\begin{table}[h]
\centering
\begin{tabular}{|c|c|c|c|c|c|c |}
 \hline
  & & mean & std & median & max & > 10  \\ \hline

  \multirow{2}{*}{frontal}  &error (mm)  & 2.12 & 4.56 & 1.00 & 38.00 & 22 \\
   
&error (slice)   & 1.53 & 4.22 & 0.67 & 45.71 & 15 \\
\hline
  \multirow{2}{*}{sagittal}  &error (mm) & 1.99 & 5.41 & 1.00 & 52.00 & 28
 \\
   
&error (slice)   & 1.41 & 5.02 & \textbf{0.50} & 65.00 & 23 \\
\hline

\end{tabular}
\caption{\label{tab:crossval_error}3-fold cross validation results for slice detection comparing the automatic slice detection with the ground truth (n=1070) with either using the frontal MIP or sagittal MIP as input. The error was obtained by computing the absolute difference in position between the automatically detected slice and the ground truth slice. The error is reported in mm and number of slices, with the latter being computed from the error in mm by dividing it by the slice thickness for a given image, without rounding. We also report the number of outlier images that have an error greater than 10 (either in mm or number of slices). 
}
\end{table}

Results reported in Table \ref{tab:interrater_error}  correspond to the error between human annotators.
The error in slices was computed by dividing the error in mm of a given image by the slice thickness, without rounding.

\begin{table}
\centering
\begin{tabular}{|c|c|c|c|c|}
\hline
   & mean & std & median & max  \\ \hline
 Error between A and B (mm)  & 1.90  & 1.76  & 1.00 & 9.00 \\

 Error between A and B (number of slices)  & 1.94 & 2.36 & 0.80 & 11.43\\
 \hline
 Error between A/B and assigned ground truth (mm)  & 1.14  & 0.97  & 1.00 & 5.00 \\

  Error between A/B and assigned ground truth (number of slices) & 0.97 & 1.18 & 0.40 & 5.71 \\
 \hline
\end{tabular}
\caption{\label{tab:interrater_error}Error for L3 slice level between annotators A and B was computed as the absolute difference between their respective slice positions (n=1070). Errors are reported in mm and in number of slices. To train the slice detection algorithm, we used as ground truth the mean slice location from both annotators. With the newly assumed ground truth, we report the error between any of the annotators and the ground truth (the error is the same for both annotators as the absolute difference with their mean is symmetric).}
\end{table}

\subsection{Segmentation}

For muscle segmentation, 3-fold cross validation Dice scores were 0.97±0.02, 0.95±0.04, 0.94±0.04 for erector spinae, psoas, and rectus abdominus, respectively, and 0.96±0.02 for the combined muscle mass. Figure 4 summarises the results.

\begin{figure}[h]
\centering
\includegraphics[width=0.8\textwidth]{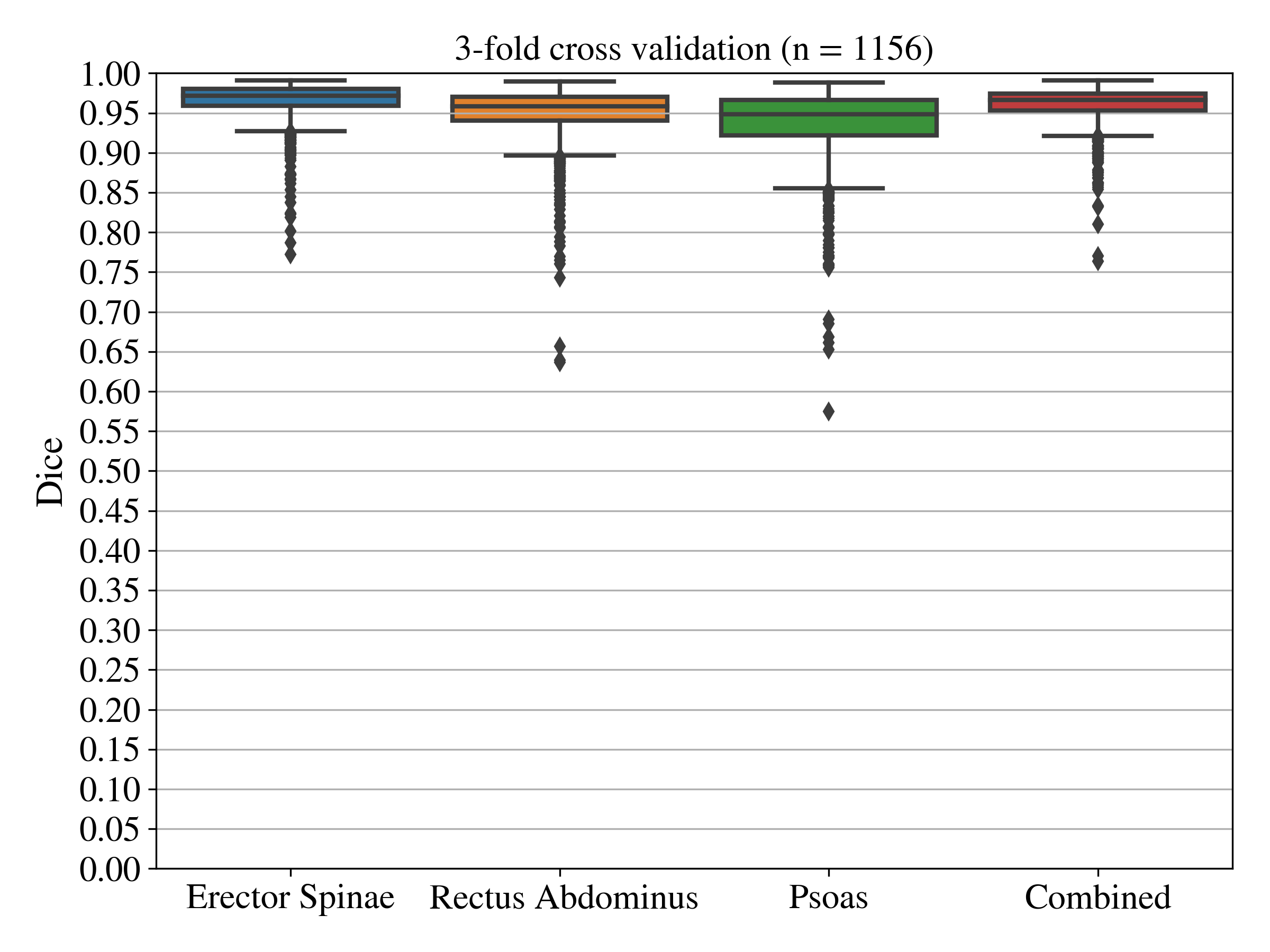}
 
\caption{\label{fig:seg_boxplot}Boxplot of Dice overlap scores between the auto-segmentation output and the ground truth for the different muscle groups as well as combined, obtained via 3-fold cross validation.  }
\end{figure}

\subsection{Sarcopenia measurements}

\begin{figure}[h]
    \centering
     \begin{subfigure}[b]{0.44\textwidth}
        \includegraphics[width=1\textwidth]{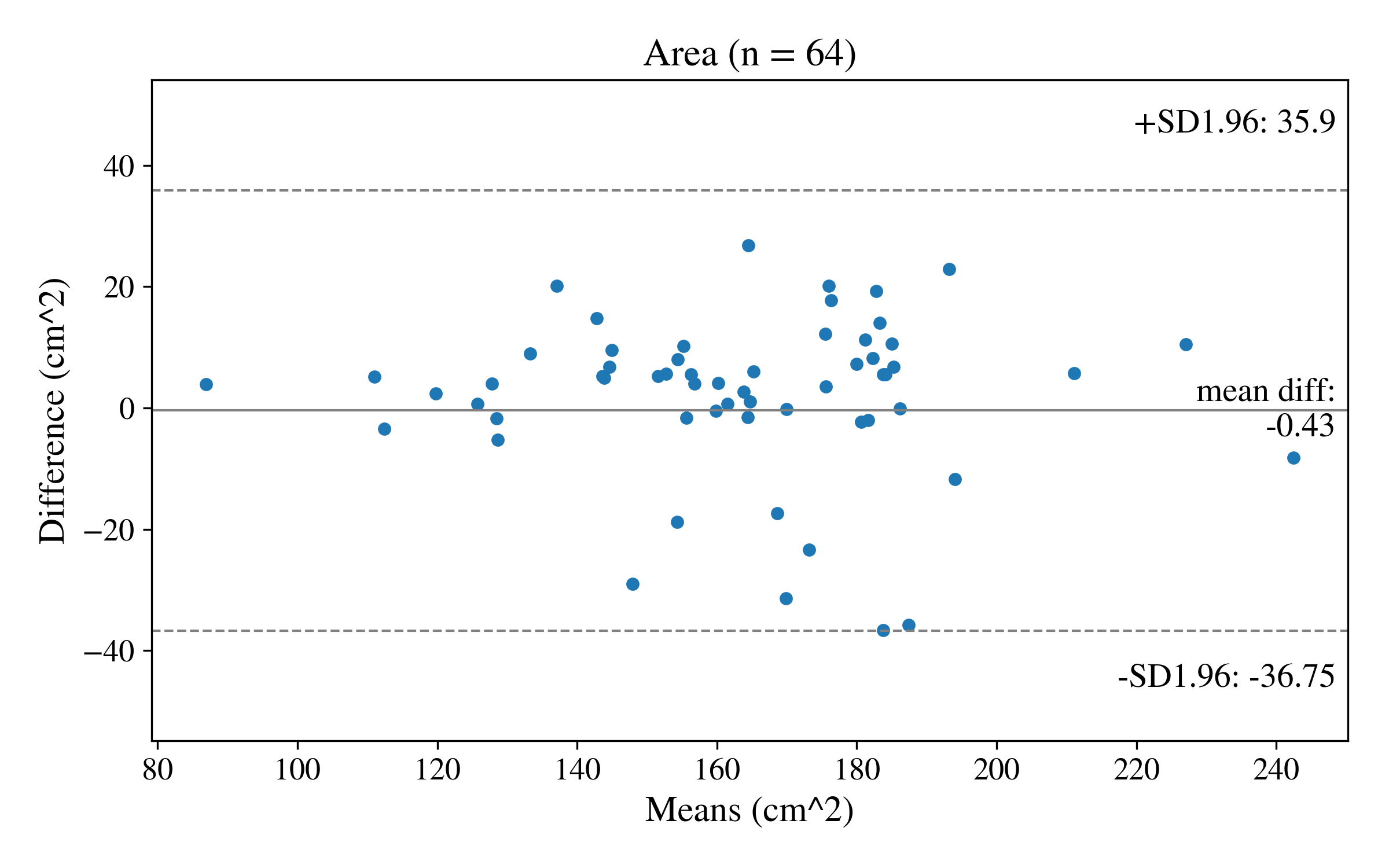}
       
         \end{subfigure}
     \begin{subfigure}[b]{0.44\textwidth}
        \includegraphics[width=1\textwidth]{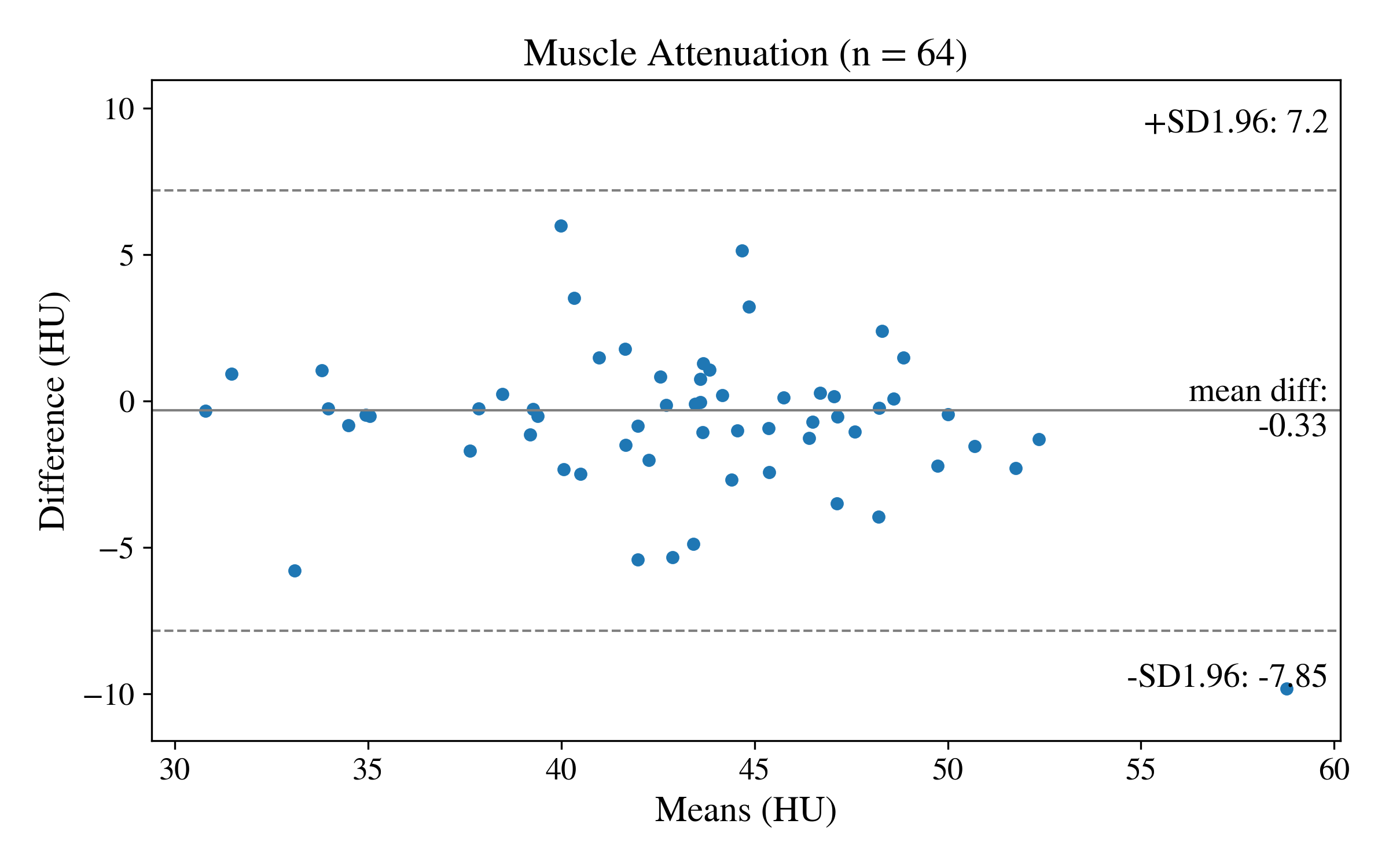}
      
         \end{subfigure}
        
\caption{\label{fig:inter_annotator_seg}Bland-Altman plots for muscle area and attenuation for the subset (n=64) of images that the two annotators independently performed slice detection and segmentation on. }
\end{figure}

\begin{figure}
    \centering
     \begin{subfigure}[b]{0.45\textwidth}
        \includegraphics[width=1\textwidth]{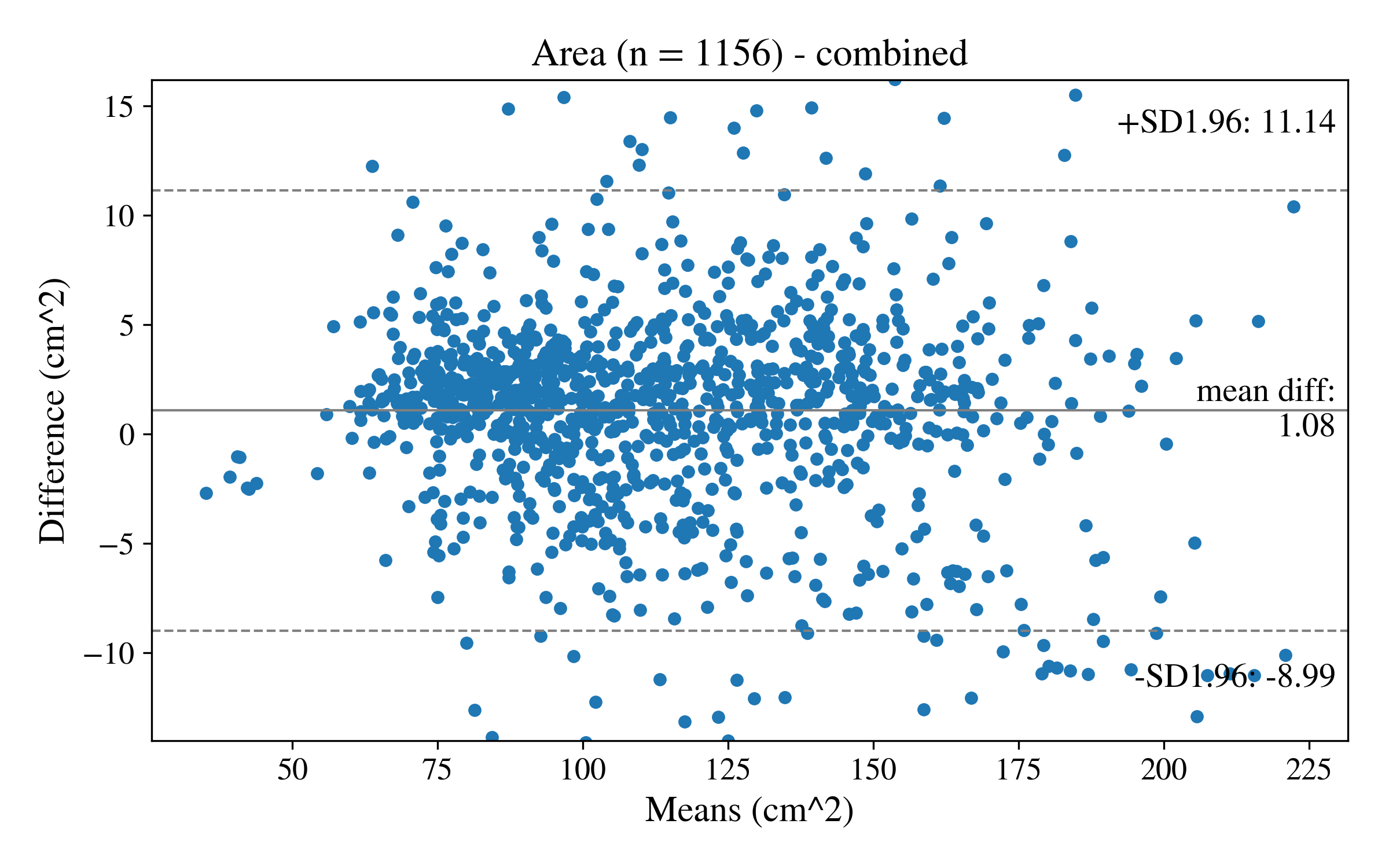}
       
         \end{subfigure}
     \begin{subfigure}[b]{0.45\textwidth}
        \includegraphics[width=1\textwidth]{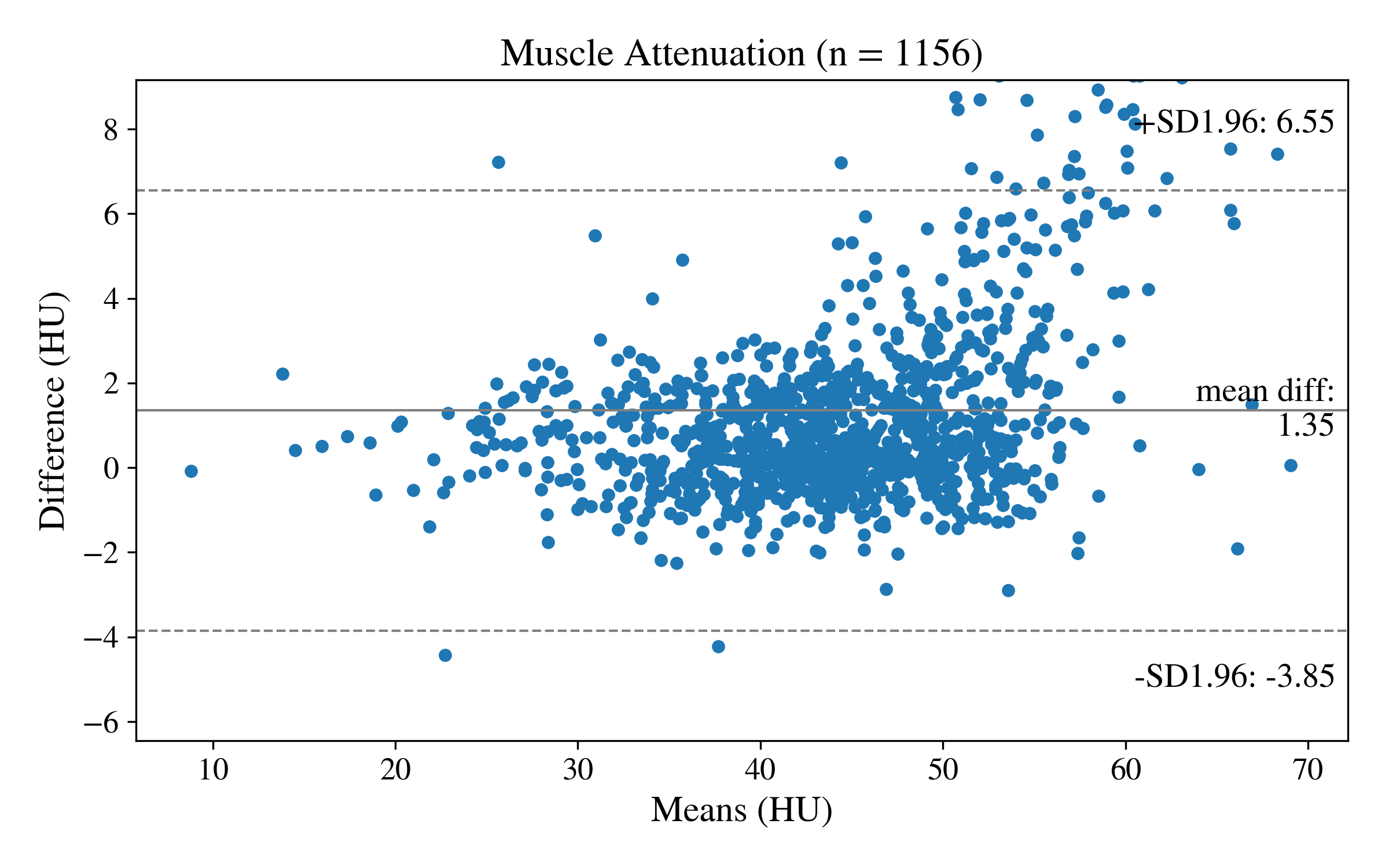}
     
         \end{subfigure}
        
\caption{\label{fig:ma_area_auto}Bland-Altman plots for muscle area and attenuation comparing the auto-segmentation output with the ground truth segmentation for the combined muscle mass.}
\end{figure}

\begin{figure}
    \centering
     \begin{subfigure}[b]{0.31\textwidth}
        \includegraphics[width=1\textwidth]{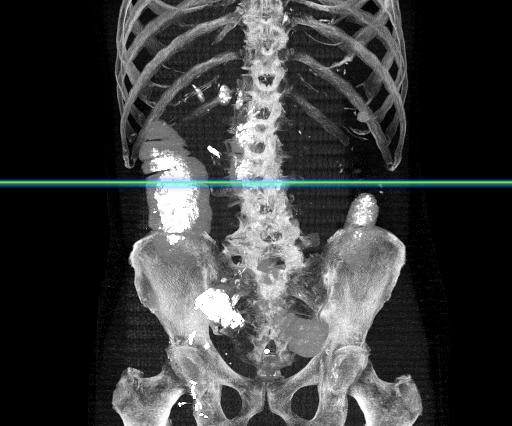}
        \caption{\\frontal}
         \end{subfigure}
     \begin{subfigure}[b]{0.31\textwidth}
        \includegraphics[width=1\textwidth]{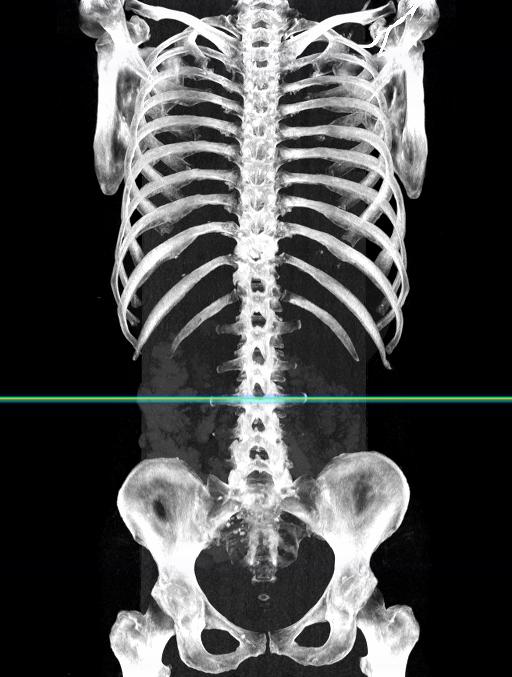}
        \caption{\\frontal}
         \end{subfigure}
            \begin{subfigure}[b]{0.31\textwidth}
        \includegraphics[width=1\textwidth]{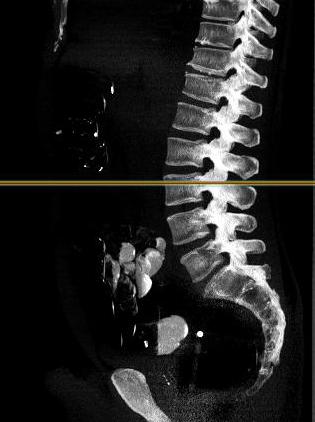}
        \caption{\\sagittal}
     
    \end{subfigure}
\caption{\label{fig:example_output}Examples of prediction output. The predicted confidence map is overlaid on the images.  The confidence map is stretched out along the x-axis for visualisation purposes.}
\end{figure}

\begin{figure}[h]
\centering
\includegraphics[width=0.50\textwidth]{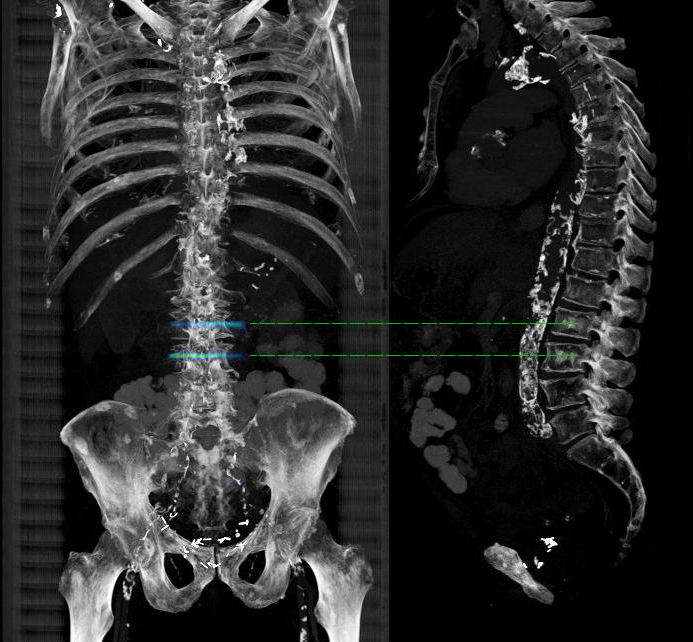}
 
\caption{\label{fig:transitional} Output on a transitional vertebra case. The confidence map indicates two potential locations for L3.   }
\end{figure}

Results in Figures \ref{fig:inter_annotator_seg} and \ref{fig:ma_area_auto} show the Bland-Altman plots for muscle attenuation and muscle area for the subset (n=64) of images that the two annotators independently performed slice detection and segmentation on, as well as the output of our proposed method. A paired t-test comparing the muscle area and attenuation computed with that of the 2nd annotator showed no statistically significant differences for muscle area (p=0.9503) and muscle attenuation (p=0.822).

\subsection{Discussion}

\begin{figure}[h]

    \centering
    
        \begin{subfigure}[t]{0.48\textwidth}
        \includegraphics[width=1\textwidth]{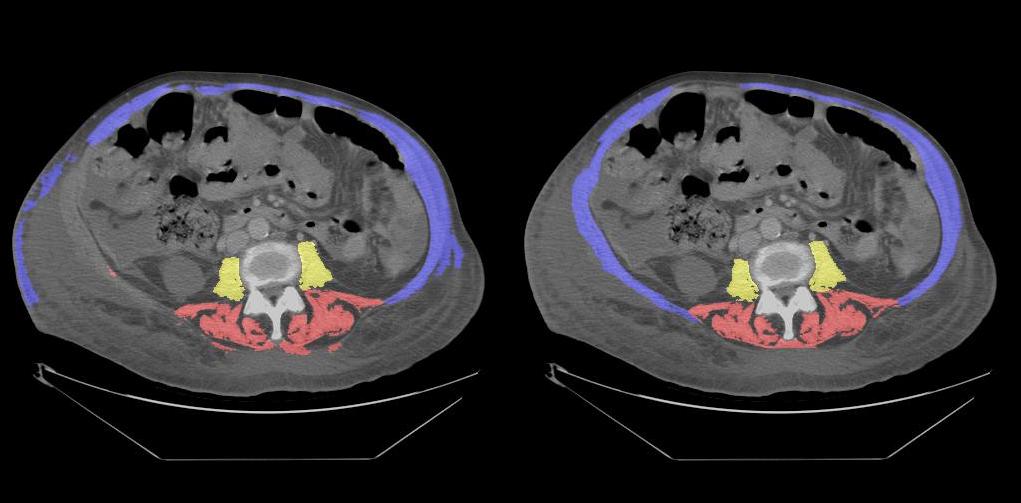}
        \caption{ An ovarian cancer patient case with auto-segmentation failure (Left). Ground truth on the right. In this challenging case, there is extensive oedema /fluid in the right subcutaneous tissue, possibly due to a drain complication, which may have contributed to the failure. \label{fig:ovarian_fail} }
    \end{subfigure}
   \quad
            \begin{subfigure}[t]{0.48\textwidth}
        \includegraphics[width=1\textwidth]{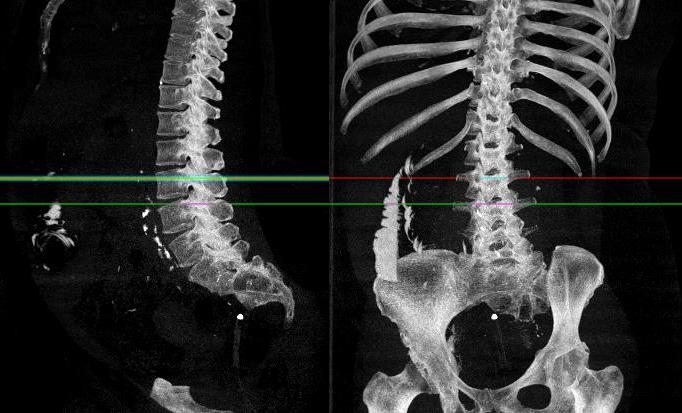}
        \caption{ Incorrect identification of the L3 level with sagittal image input. The spine is curved, resulting in an apparent merging of L5 with the sacrum. Green line is ground truth, red line is prediction.}

    \end{subfigure}
\caption{\label{fig:outliers} Examples of outlier cases.}
\end{figure}

In this study, we trained FCNNs for slice detection and muscle segmentation for the purpose of using them as a tool for automatic single-slice sarcopenia assessment.
Results show that slice detection performance is on par with human level performance, with a median error equal to 1mm, similar to the human annotator error.
There still remains a small subset of images where the prediction is maximally off by one vertebra (outliers are images with an error more than 10mm). An inspection of the outlier images revealed that roughly half of them have an apparent reason that might explain why the network made an incorrect prediction. 
 Figure \ref{fig:outliers} shows such an example of outliers.

It is potentially likely, though, despite the best efforts of the annotators that some transitional vertebrae cases have included into the training set, given that the cases are prevalent in the general population at 15-3\%, while only 57 out of 1070, which amounts to about 5\% were found in our dataset. With transitional vertebrae cases, it is not possible to determine the correct L3 level without a full view of the spine, which would allow counting from the cervical segment. We tested our network on the excluded set of transitional vertebrae images, the outputted confidence maps always resulted in predictions for one or the other potential L3 candidate, and occasionally both.

Results also showed that the auto-segmentation achieved high Dice overlap for the three muscle groups as well as combined. No statistically significant difference was observed when comparing sarcopenia measurements obtained with the proposed method to those obtained by a 2nd annotator. Nonetheless, there are a few failure cases; Figure \ref{fig:ovarian_fail} shows an example of segmentation failure on an ovarian cancer patient. 

One major limitation of this study is the lack of patient outcomes, making it difficult to assess how much of an impact offsets from L3 and differences in segmentation have on the final stratification of patients. While the results achieved in this study are promising, it remains to be seen how well the method performs on a much larger and more diverse cohort of patients.

While we have picked the L3 slice in this study, our method is flexible in that it could generalise to work on different slices at any level.  As there is still no universally accepted standard for single-slice CT sarcopenia assessment, using a generalised version of our tool would allow fast computation (within a second) of multiple different measurements in order to evaluate their diagnostic impact for sarcopenia assessment. Our code will be made publicly available for reproducibility.

\section*{Acknowledgements}
This work was supported by United Kingdom NIHR Biomedical Research Centre award to Imperial College London. We acknowledge programmatic support from Imperial College Experimental Cancer Medicines Centres (C37/A7283) and United Kingdom Medical Research Council (MC-A652-5PY80).

\bibliographystyle{plain}
\bibliography{biblio}

\end{document}